\newcommand{\Lsun} {L$_\odot$}
\newcommand{\Msun} {M$_\odot$}

\documentclass{aa}  
\usepackage{graphicx}
\usepackage{txfonts}
\begin{document}
   \title{Dust in starburst nuclei and ULIRGs:}

   \subtitle{SED models for observers}

\author {R.~Siebenmorgen\inst{1} \and E.~Kr\"ugel\inst{2}}

\offprints{rsiebenm@eso.org}

\institute{
        European Southern Observatory, Karl-Schwarzschildstr. 2,
        D-85748 Garching b. M\"unchen, Germany
\and
        Max-Planck-Institut f\"ur Radioastronomie, Auf dem H\"ugel 69,
        Postfach 2024, D-53010 Bonn, Germany }

   \date{Received May 26, 2006; accepted xxx,xxx, 200x}

  \abstract
  % context heading (optional)
 {} 
%% aims heading (mandatory) 
{We provide a library of some 7000 SEDs for the nuclei of starburst and ultra
luminous galaxies.  Its purpose is to quickly obtain estimates of the basic
parameters, such as luminosity, size and dust or gas mass and to predict the
flux at yet unobserved wavelengths.  The procedure is simple and consists of
finding an element in the library that matches the observations.  The objects
may be in the local universe or at high $z$.}
%% methods heading (mandatory) 
{We calculate the radiative transfer in spherical symmetry for a stellar cluster
permeated by an interstellar medium with standard (Milky Way) dust properties.
The cluster contains two stellar populations: old bulge stars and OB stars.
Because the latter are young, a certain fraction of them will be embedded in
compact clouds which constitute hot spots that determine the MIR fluxes.  }
%% results heading (mandatory) 
{We present SEDs for a broad range of luminosities, sizes and obscurations.  We
argue that the assumption of spherical symmetry and the neglect of clumpiness of
the medium are not severe shortcomings for computing the dust emission.  The
validity of the approach is demonstrated by matching the SED of seven of the
best studied galaxies, including M82 and Arp220, by library elements.  In all
cases, one finds an element which fits the observed SED very well, and the
parameters defining the element are in full accord with what is known about the
galaxy from detailed studies.  We also compare our method of computing SEDs with
other techniques described in the literature.}
%% conclusions heading (optional)
{}

\maketitle
%
%________________________________________________________________

\section{Introduction}

By definition, the rapid conversion of a large amount of gas into predominantly
massive ($> 8 M_\odot$) stars, or the result of such a conversion, is called a
starburst.  Starburst galaxies constitute a unique class of extragalactic
objects.  The phenomenon is of fundamental importance to the state and evolution
of the universe, as outlined, for example, in the review by Heckman (1998).
According to him, in the local universe starbursts are responsible for about
25\% of the high--mass star formation rate and for 10\% of the total luminosity;
in the early universe, beyond $z \sim 0.7$, IR luminous galaxies dominate the
star forming activity (Flo'ch et al.~2005).  Starbursts are also cosmologically
significant if one interprets the high bolometric luminosities of high redshift
galaxies to be due to star formation (Hirashita et al. 2003) at a rate so high
that it can only be maintained over a cosmologically short spell ($< 10^8$ yr).
The study of nearby starbursts would then help us to understand the processes
underlying the star formation history of the universe.

\noindent
Starbursts, we think, are triggered by the gravitational interaction
between galaxies (Kennicutt et al. 1987), but they occur, as a result
of mass and angular momentum transfer, predominantly in their nuclei,
at the center of a massive and dynamically relaxed cluster of old
stars (the bulge).  Although the region where OB stars form is
relatively small (a few hundred parsec), its luminosity often exceeds
that of the host galaxy. Starbursts are almost pure infrared objects,
opaque to stellar photons.  Whereas, on average, in the local universe
$\sim$60\% of the star formation is obscured by dust (Takeuchi et al.,
2006), in starbursts the fraction is typically 90\%.  To interpret
infrared observations and to arrive at a self--consistent picture for
the spatial distribution of stars and interstellar matter in the
starburst nucleus and of the range of dust temperatures, one has to
simulate the transfer of continuum radiation in a dusty medium.  Line
emission is energetically negligible.

\noindent
A starburst has four basic parameters: total luminosity, $L$, dust or
gas mass, $M_{\rm d}$ or $M_{\rm gas}$, visual extinction, $A_{\rm
V}$, and size.  Size, $A_{\rm V}$ and $M_{\rm d}$ are, of course,
related, for a homogeneous density model, only two of them are
independent. The luminosity follows observationally in a straight
forward way by integrating the spectral energy distribution over
frequency, $M_{\rm d}$ is readily derived from a millimeter continuum
data point, if available, and the outcome is almost independent of the
internal structure or viewing angle of the starburst.  The size is
best obtained from radio observations as it does not suffer extinction
by dust.  It does not matter whether the radio emission is thermal or
non-thermal, both are connected to high--mass stars.

\noindent
In this paper, we present a set of SEDs for starbursts covering a wide
range of parameters.  Anyone with infrared data and interested in
their interpretation can compare them with our models, find an SED
that matches (after normalization of the distance) and thus constrain
the properties of the starburst under investigation without having to
perform a radiative transfer computation himself.

\begin{figure*} [htb]
\includegraphics[width=18cm,height=19.3cm,angle=0]{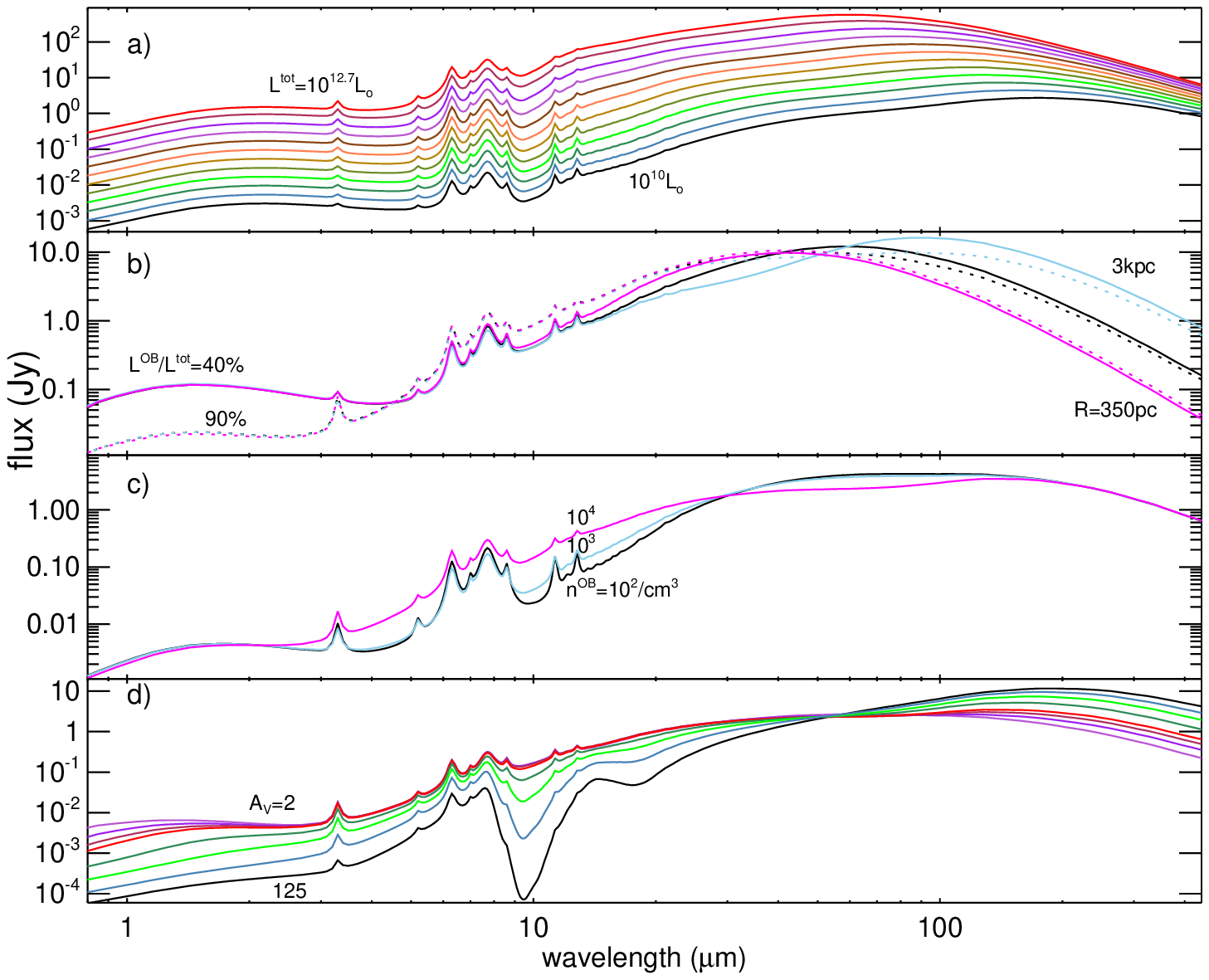}
\caption{Influence of starburst parameters on the SED for a distance
of 50 Mpc.  The parameters which are kept constant are listed in
square parentheses. \ {\bf a)} Total luminosity is varied between
$L^{\rm {tot}} = 10^{10}$ and $10^{12.7}$\Lsun; [$R=3$kpc, $A_{\rm V}
\sim 17$mag, $L_{\rm {OB}} / L^{\rm {tot}} = 0.6$ and $n^{\rm hs} =
10^3$ cm$^{-3}$].  {\bf b)} Here we vary two parameters: the radius of
the nucleus from $R=0.35$ over 1 to 3kpc, and the luminosity ratio:
$L_{\rm {OB}} / L^{\rm {tot}} = 0.4$ (full lines), 0.9 (dotted);
[$L^{\rm {tot}} = 10^{11.1}$\Lsun, $A_{\rm V} \sim 4.5$mag and $n^{\rm
hs} = 10^4$ cm$^{-3}$].  {\bf c)} Variation of the hot spot density:
$n^{\rm hs} = 10^2$, $10^3$ and $10^4$ cm$^{-3}$; [$L^{\rm {tot}} =
10^{10.5}$\Lsun, $R=3$kpc, $A_{\rm V} \sim 9$mag, $L_{\rm {OB}} /
L^{\rm {tot}} = 0.9$].  {\bf d)} Variation of the dust extinction:
$A_{\rm V} \sim $2.2, 4.5, 6.7, 9, 18, 35, 70 and 125mag; [$L^{\rm
{tot}} = 10^{10.5}$\Lsun, $R=3$kpc, $L_{\rm {OB}} / L^{\rm {tot}} =
0.9$ and $n^{\rm hs} = 10^4$cm$^{-3}$]. \label{para4.ps}}
\end{figure*}

\section{Dust model and radiative transfer}

\noindent 
A description of the dust model and the radiative transfer can be found in
chapter 12 and 13 of Kr\"ugel (2003), we here only summarize the salient points.
We use standard dust.  It consists of silicate and amorphous carbon grains with
an MRN size distribution ($n(a)\propto a^{-3.5}$, \ $a \sim 300 \ldots
2400$\AA), and a population of small graphite grains ($a \sim 10 \ldots
100$\AA).  There are also two kinds of PAHs ($N_{\rm C}=38, N_{\rm H}=12$ and
$N_{\rm C}=250, N_{\rm H}=48$, where $N_{\rm C}$, $N_{\rm H}$ are the number of
C, H atoms, respectively).  By mass, 63\% of the dust is in silicates, 37\% in
carbon of which 60\% is amorphous, 38\% graphitic and 2\% in PAHs.  About 5\% of
the graphitic particles are so small ($< 60$\AA) that their temperature
fluctuates.  Such a dust mixture produces reddening in rough agreement with the
standard interstellar extinction curve for $R_{\rm V} = 3.1$.

\noindent 
An important feature of our model is the division of the sources in the
starburst nucleus into two classes.

{\it a)} OB stars in dense clouds with total luminosity $L_{\rm OB}$.
The immediate surroundings of such a star constitutes a {\it hot spot}
and its emission must be evaluated separately as they, more than
anything else, determine the MIR part of the SED of a galactic nucleus
(Kr\"ugel \& Tutokov 1978, Kr\"ugel \& Siebenmorgen 1994).  The outer
radius of a hot spot, $R_{\rm hs}$ is given by the condition of equal
heating of the dust from the star and from the ambient radiation
field.  The hot spots, whose total volume is small compared to the
volume of the galactic nucleus, are presented in the radiative
transfer equation by a continuously distributed source term
$\varepsilon^{\rm hs}_\nu(r)$, where $r$ is the distance towards the
center of the galactic radius.  For a fixed OB stellar luminosity,
$\varepsilon^{\rm hs}_\nu$ is sensitive to the assumed density in the
hot spot, $\rho^{\rm hs}$.

{\it b)} All other stars of total luminosity $L_{\rm tot}-L_{\rm OB}$.  These
are mainly the old bulge stars of low brightness and surface temperature, but
also hotter stars not enveloped in a dense cloud.  This population is presented
in the radiative transfer equation by a continuously distributed source term
$\varepsilon^{\rm bulge}_\nu(r)$. 

\noindent 
The model galactic nucleus is a sphere (of radius $R$) and the radiative
transfer is computed with ray tracing.  The intensity, $I_\nu(p,z)$, is a
function of frequency $\nu$, impact parameter $p$, and coordinate $z$.  At
different $\nu$ and $p$, we solve along the $z$--axis the equations
\begin{equation}\label{RadTr_int1} 
I^+(\tau) \ = \ 
I^ +(0) \ \/ \ e^{-\tau} \ + \ \int_0^\tau S(x) \,e^{x-\tau}\, dx 
\end{equation} 
\begin{equation}\label{RadTr_int1a} 
I^-(t) \ = \ \int_0^t S(x) \,e^{x-t} \, dx
\end{equation}

\noindent 
$I^+$ and $I^-$ refer to the plus and minus direction of $z$, respectively.  The
indices $p$ and $\nu$ have been omitted.  The optical depth $\tau$ is zero at
$z=0$ and increases with $z$, the optical depth $t$ is zero at the edge of the
nucleus (where $z_{\rm e} =\sqrt{R^2-p^2}$) and decreases with $z$.  There is no
radiation incident from outside, so $I^-(z=z_{\rm e})=I^-(t=0)=0$, and symmetry
requires $I^+=I^-$ at $z=0$.  The source function (dropping sums over different
kinds of dust particles) equals

\begin{equation}\label{RadTr} 
S_\nu = \frac{1} { K^{\rm ext}_\nu} \cdot \left[ \varepsilon^{\rm
hs}_\nu + \varepsilon^{\rm bulge}_\nu + K^{\rm sca}_\nu J_\nu + K^{\rm
abs}_\nu \displaystyle{\int P(T) B_\nu(T)\, dT} \right] 
\end{equation}

\noindent 
The term $K^{\rm abs}_\nu \int P(T) B_\nu(T)$ describes the emission
of dust grains.  If they are big, the probability density $P(T)$
equals the $\delta$--function $\delta(T_{\rm d})$ where $T_{\rm d}$
follows from the equilibrium between radiative heating and cooling.
For small grains, $P(T)$ is evaluated in an iterative scheme similar
to the method of Guhathakurta \& Draine (1989).  $J_\nu$ is the
galactic radiation field.  As we assume isotropic scattering, we
reduce the Mie scattering efficiency, $Q^{\rm sca}$, by the factor
$(1-g_\nu)$, where $g_\nu$ is the asymmetry factor.  All quantities
depend on the galactic radius, $r$.  The emission from the hot spots
is calculated separately in a radiative transfer program for an OB
star in a spherical cloud of density $\rho^{\rm hs}$ bathed in the
galactic radiation field.

\begin{figure*}
\center \includegraphics[angle=90,width=13.6cm]{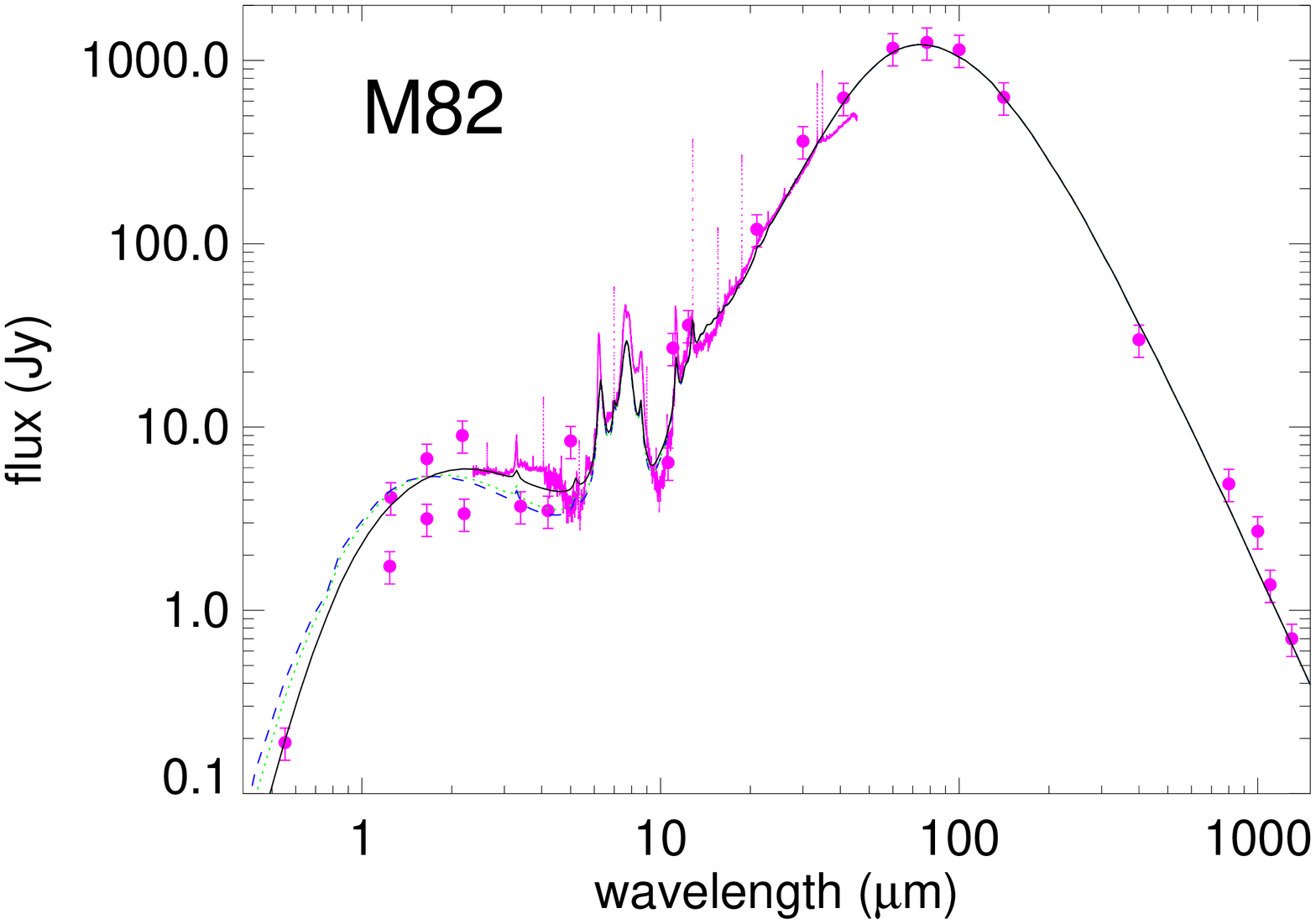}
\vspace{-1.cm}
\center \includegraphics[angle=90,width=13.6cm]{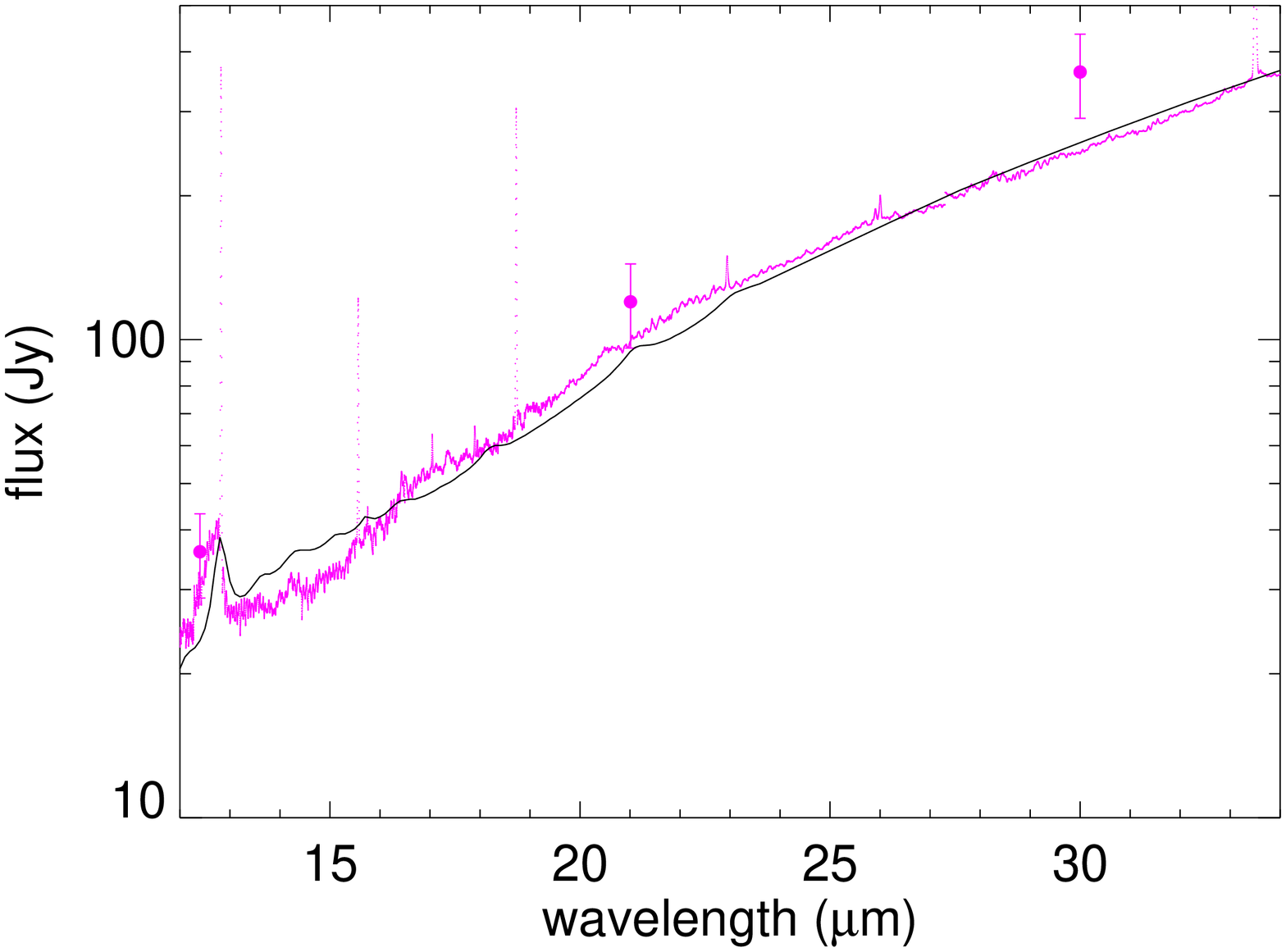}
\caption{SED of the central region of M82, data points with $1\sigma$
 error bar.  Full line: library model with parameters in Tab.~1.  To
 fit the data below 5$\mu$m, we added to the SED library spectrum a
 blackbody, either unreddened ($T= 2500$ K, full line), or reddened
 ($T=8000$K, $A_{\rm V} =4$mag, dashed, or $T=5000$K, $A_{\rm V}
 =3$mag, dotted).  Full $0.4 - 1500\mu$m wavelength range (top), a
 zoom into the $12 - 34\mu$m region (bottom).  Data references
 (1300$\mu$m: Kr\"ugel et al. (1990); 1100 and 800$\mu$m: Hughes et
 al. (1990); 400$\mu$m: Jaffee et al. (1984); FIR: Telesco \& Harper
 (1980), Rieke \& Low (1972), Rieke et al. (1980), Telesco \& Gezari
 (1992); IRAS; NIR photometry in 40$''$ -- 100$''$ aperture: Kleinmann
 \& Low (1970), Jarret et al. (2003), Aaronson (1977) and Johnson
 (1966); between 2.3--40.4$\mu$m ISOSWS spectrum: Sloan et
 al. (2003)). \label{M82.ps}}
\end{figure*}

\section{Parameter space of the model grid}

\noindent
For our set of SEDs, we vary in the calculations the following five parameters:

\begin{enumerate} 
\item total luminosity $L^{\rm {tot}}$ from $10^{10}$ to
$10^{14}$\Lsun \/ in steps of $0.1$ in the exponent, 

\item  nuclear radius, $R= 0.35$, 1 and 3 kpc,

\item the visual extinction from the edge to the center of the nucleus: 
$A_{\rm V} \simeq 2.2, 4.5, 7, 9, 18, 35, 70$ \/ and 120 mag,

\item ratio of the luminosity of OB stars with hot spots to the total
luminosity: $L_{\rm {OB}} / L^{\rm {tot}} = 0.4, 0.6$ and 0.9,

\item dust density in the hot spots. For a gas--to--dust ratio of 150, the 
corresponding hydrogen number densities are $n^{\rm {hs}} = 10^2,
10^3$ and $10^4$ cm$^{-3}$.  The density is constant within the hot
spot.
\end{enumerate}
 
\noindent
Not all parameter combinations are included in the set of SEDs because some are
astronomically unlikely (for instance, very high $L^{\rm {tot}}$ and very little
extinction).  Altogether, the grid contains 7000 entries.

\noindent
The dust density in the nucleus, $\rho$, is spatially constant,
$\partial\rho /\partial r =0$.  Its value follows from the extinction
$A_{\rm V}$ and the nuclear radius, $R$.  The dust mass, $M_{\rm d}$,
is then given by $4\pi\rho R^3/3$ and increases linearly with $A_{\rm
V}$.  For example, for $R=350$pc and $A_{\rm V}=18$, the gas mass,
$M_{\rm gas}$, is $1.7\times 10^{8}$\Msun. The density of all stars is
centrally peaked, $\rho_*(r) \propto r^{-1.5}$.  The OB stars are
always confined to the inner 350 pc, and they have a fixed luminosity
and surface temperature ($2\times 10^4$ L$_\odot$, \ $T_{\rm eff}=
25000$ K). The bulge stars fill the total volume.  As they do not form
hot spots, we need not specify the luminosity of a single star.  For
$L_{\rm tot} \leq 10^{12.7}$\Lsun, their surface temperature, $T_{\rm
eff}$, equals 4000 K (old giants).  When $L_{\rm tot}
>10^{12.7}$\Lsun, we assume $T_{\rm eff}$ = 25000 K, which means that
they consist mainly of OB stars, but outside compact clouds.

\noindent
Fig.~\ref{para4.ps} illustrates the changes in the SED when one parameter is
varied while all others stay fixed; fluxes refer to a source distance of 50
Mpc. Panel a) informs us how a rise in luminosity shifts the far IR peak to
shorter wavelengths.  The flux then increases in the near IR much more strongly
than at submillimeter wavelengths.  If $L_{\rm tot} \ge 10^{12.5}$\Lsun \/, the large
grains become so warm that at $\lambda > 11\mu$m they outshine the PAH features.
In panel b), we see that as the source becomes bigger, the dust gets cooler
(maximum emission at longer wavelengths).  This is not because the dust is then,
on average, farther away from the source, but because the dust mass, $M_{\rm
d}$, grows with $R^2$ when $A_{\rm V}$ is constant, and the mean dust
temperature is determined by $L^{\rm {tot}}/ M_{\rm d}$.  We also see that a
high ratio of $L_{\rm {OB}}/L^{\rm {tot}}$ enhances the near IR flux.  Panel c)
shows the influence of the density in the hot spots for $n^{\rm {hs}} = 10^2$, $10^3$ and
$10^4$ cm$^{-3}$, it is particular strong in the MIR.  Panel d) depicts the
influence of the optical depth.  Large values suppress the near IR emission and
produce absorption in the 10$\mu$m, and for very high extinctions ($A_{\rm
V}\geq 70$mag) also in the 18$\mu$m silicate bands.

\begin{figure*}
   \centering
\hbox{   
\vspace{0cm}
\hspace{-0.cm}
\includegraphics[angle=90,width=9cm,height=10.5cm]{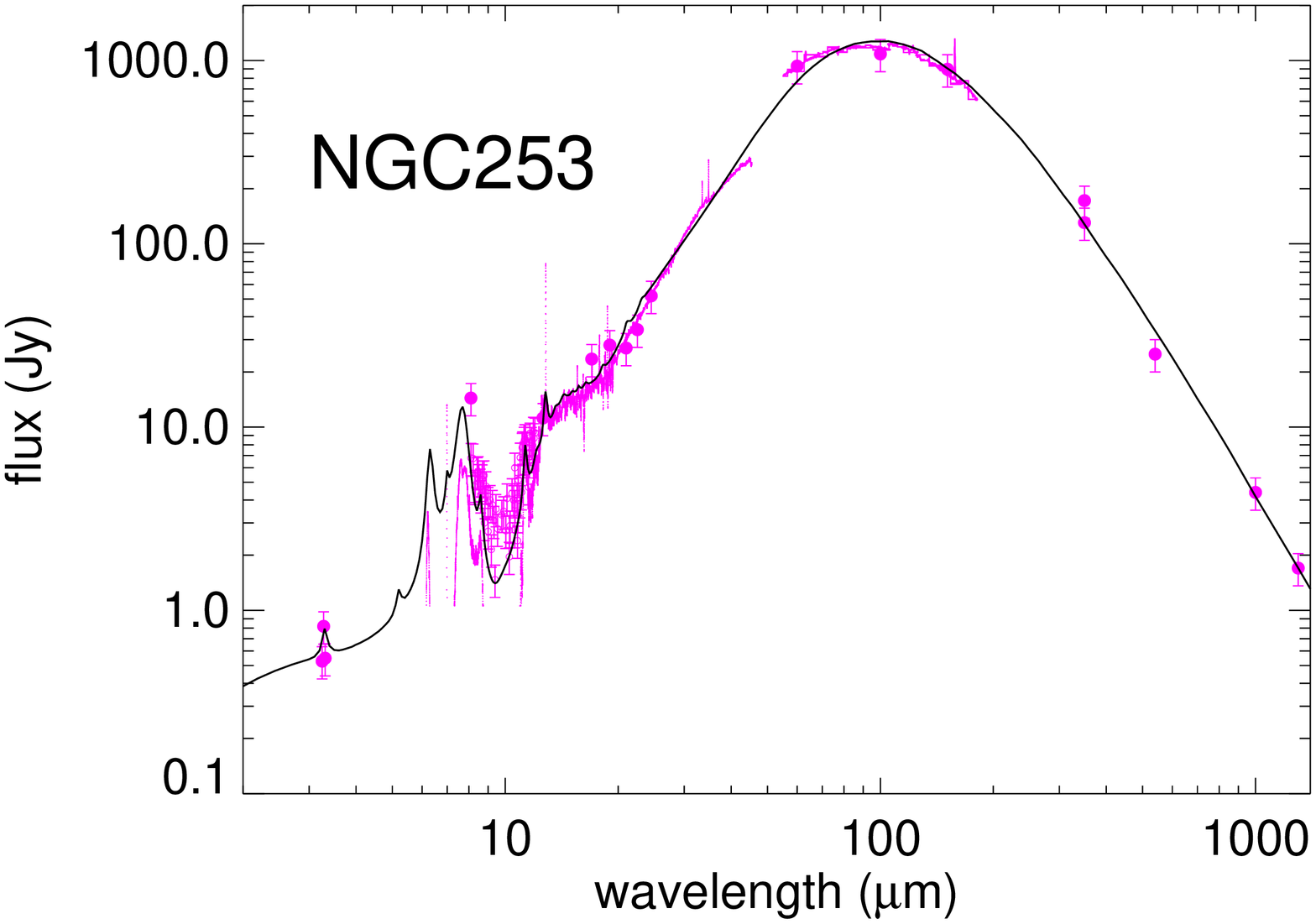}
\vspace{0cm}
\hspace{-1.cm}
\includegraphics[angle=90,width=9cm,height=10.5cm]{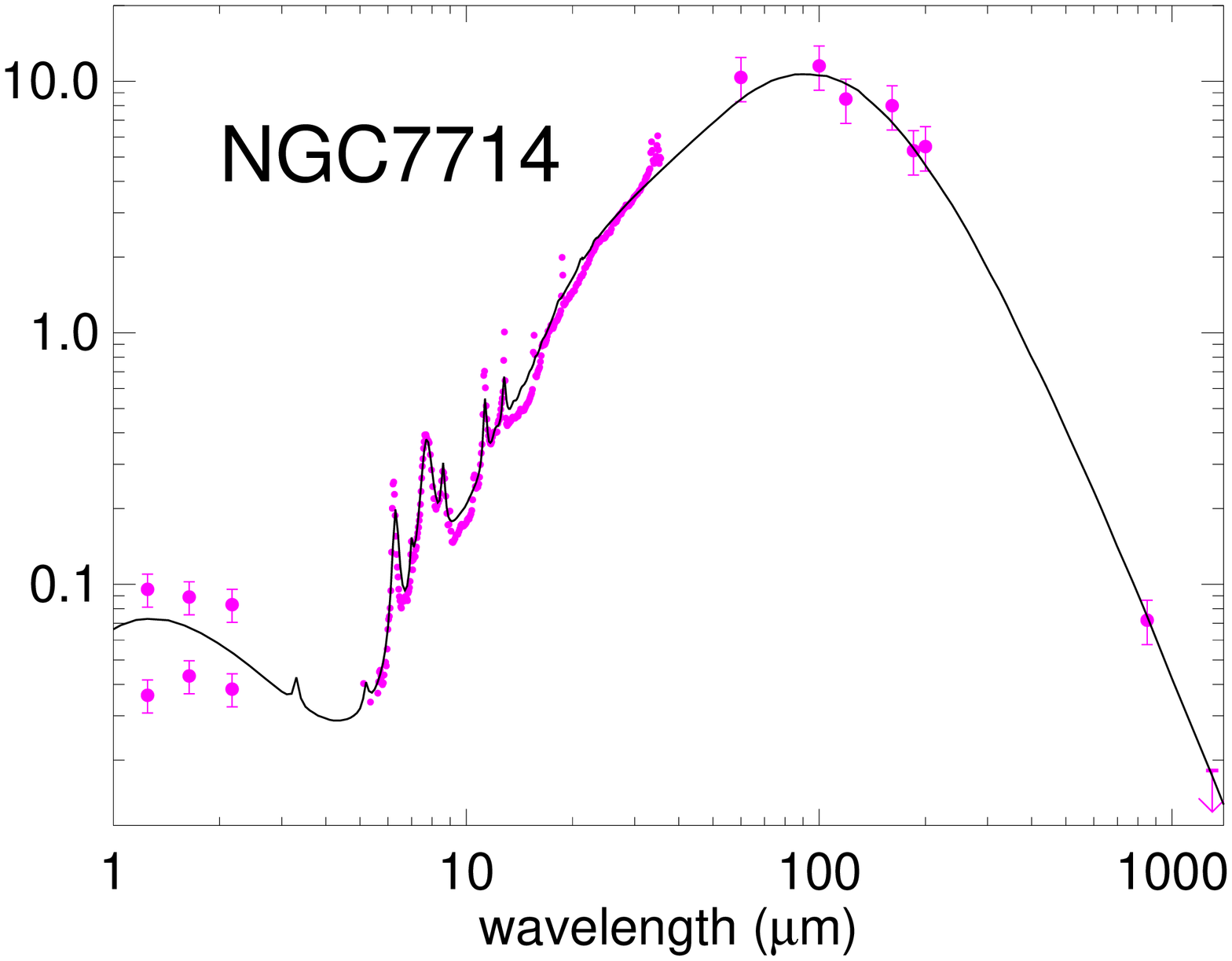}}
 \caption{SEDs of NGC253 and NGC7714, data points with $1\sigma$
error bar.  Models: full lines, model parameters in Tab.~1.  Data for
NGC253 (ISOSWS: Sloan et al. (2003); ISOLWS and ISOPHT: Radovich et
al. (2001); NIR: Rieke \& Low (1975); IRAS; submm: Rieke et
al. (1973), Hildebrand et al. (1977), Chini et al. (1984)).  Data for
NGC7714 (NIR: Spinoglio et al. (1995), Jarret et al. (2003); Spitzer
IRS: Brandl et al. (2004); IRAS; ISOPHT: Kr\"ugel et al. (1998);
850$\mu$m: Dune et al. (2000); 1.3mm: Kr\"ugel et al. (1998)).  
\label{sed1.ps}}
\end{figure*}

   \begin{figure*} %\hspace{-1.cm}
\vspace{0cm}
\hspace{-1.cm}
   \includegraphics[angle=90,width=19.cm,height=13.cm]{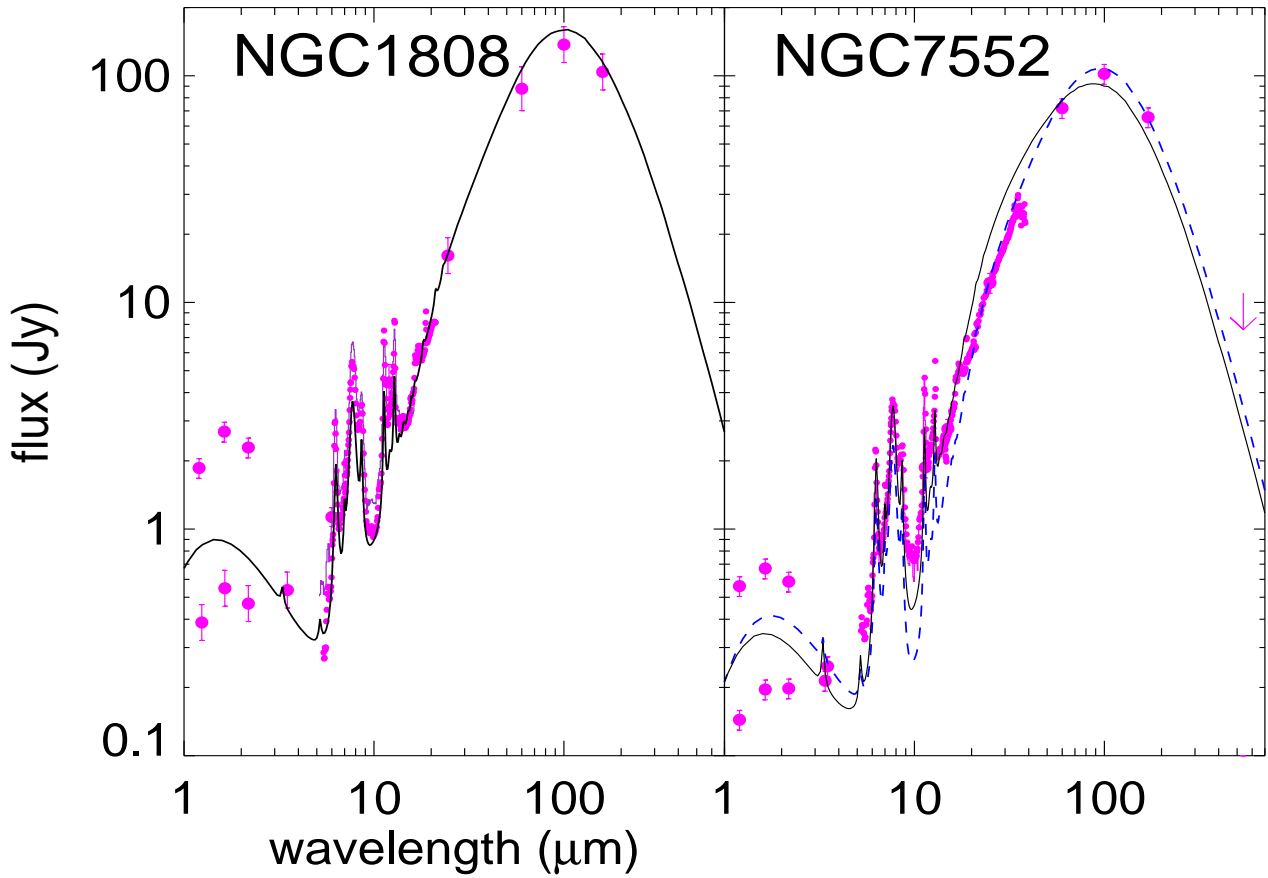}
   \caption{SEDs of NGC1808 and NGC7552, data points with $1\sigma$
error bar.  Models: full and dashed lines, model parameters in Tab.~1.
Data for NGC1808 (NIR: Glass (1976), Jarret et al. (2003); IRAS;
ISOPHT 160$\mu$m and ISOCAM spectroscopy: Siebenmorgen et
al. (2001). Data for NGC7552 (NIR: Glass (1976), Jarret et al. (2003);
ISOCAM: Roussel et al. (2001); TIMMI2: Siebenmorgen et al. (2004);
Spitzer IRS of nucleus: Kennicutt et al. (2003); IRAS; submm: Stickel
et al. (2004), Hildebrand et al. (1977)).  
\label{sed2.ps}}
\end{figure*}

   \begin{figure*}
\hspace{-1.cm}
   \includegraphics[angle=90,width=19.cm,height=13.cm]{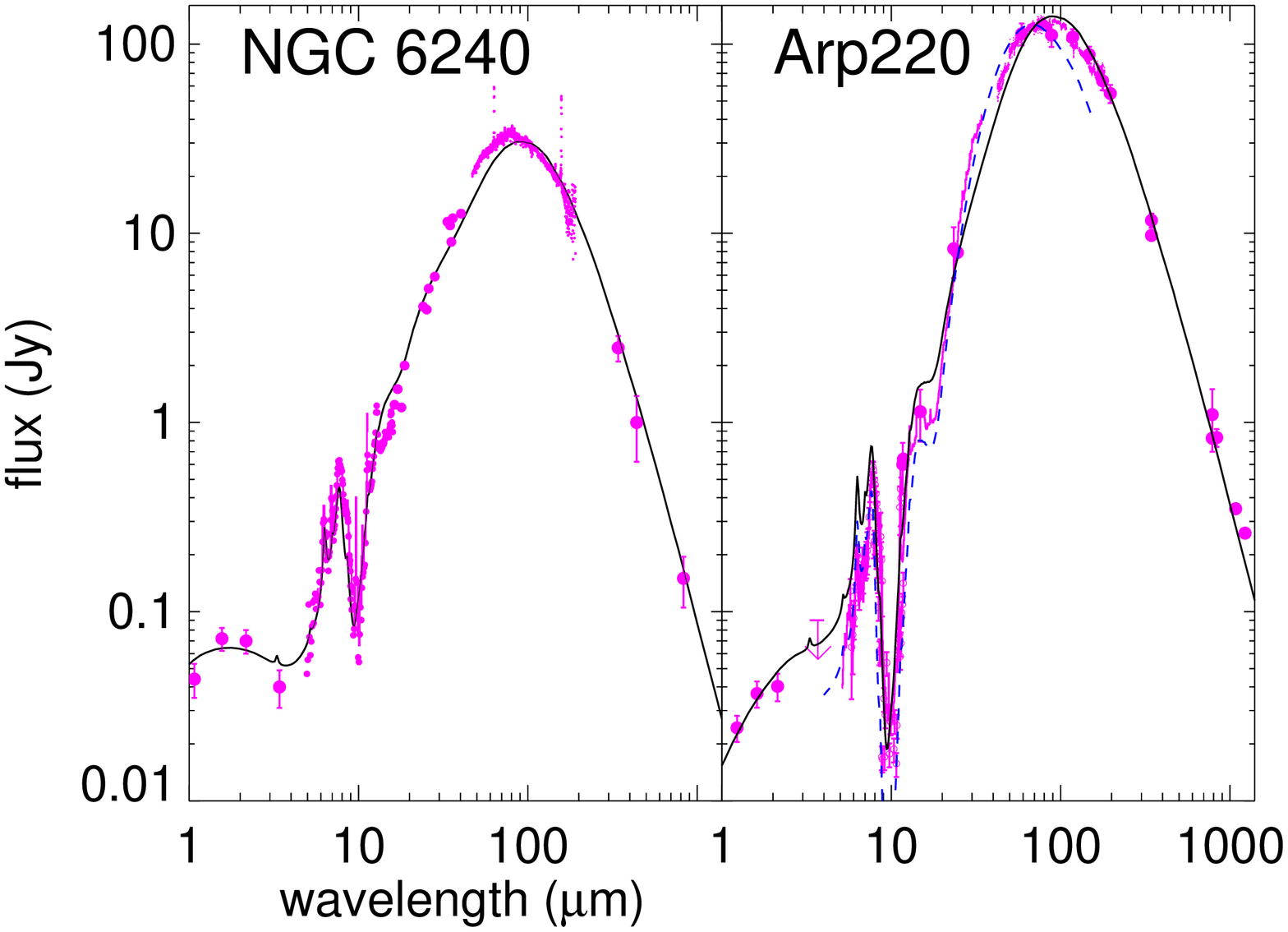}
   \caption{SEDs of NGC6240 and Arp220, data points with $1\sigma$ error bar.
Models: full and dashed lines, model parameters in Tab.~1.
To match JHK photometry of NGC6240, we added a 4000K black body to the starburst
   model.
Data for NGC6240 (NIR: Spinoglio et al. (1995); ISOPHT
   and submm: Klaas et al. (1997, 2001); 350$\mu$m: Benford (1999);
   ISOCAM spectroscopy:  (Laurent et al. 2000); ISOPHT and ISOSWS: Lutz
   et al. (2003)).
Data for Arp220 (2MASS: Jarret et al. (2003); IRAS;
   ISOPHT: Klaas et al. (2001); ISOCAM: Siebenmorgen \& Efstathiou
   (2001); submm: Benford (1999), Rigopoulou (1996), Dunne et
   al. (2000), Carico et al. (1992), Chini et al. (1986); ISOLWS
   archive spectrum is scaled to match the ISOPHT photometry and
   Spitzer IRS spectrum (this work)).  
\label{ulirg.ps}} 
\end{figure*}

\section{Testing the SED library}

\subsection{Fitting prototypical galaxies}

\noindent 
We put our library of $\sim$7000 theoretical SEDs to the test by applying it to
seven famous and well studied galaxies of the local universe, five starbursts
(M82, NGC253, NGC7714, NGC1808, NGC7552) and two ULIRGs (NGC6240 and Arp220).
The observational data and our fits are displayed in Fig.\ref{M82.ps} to
Fig.\ref{ulirg.ps}, underlying model parameters are listed in Tab.~1.  The seven
galaxies, discussed in more detail below, cover a wide range of luminosities and
we check:

\begin{itemize}
\item  whether their observed spectra can be reasonably matched or, at
least, bracketed by elements of the set;

\item  whether there is only one matching element or, at most, a few which
are similar in their basic parameters;

\item whether the parameters of the matching element are meaningful,
i.e.~whether they are consistent with the information about the structure of the
galactic nucleus which we already have. \\
\end{itemize}

{\bf {\it M82}}
\smallskip

\noindent 
The present model for this archetype starburst is similar to the one proposed
before (Kr\"ugel \& Siebenmorgen 1994).  The latter was shown only for $\lambda
\ge 3\mu$m.  At shorter wavelengths, the observed flux does not steeply decline,
as the old model predicts and as one would expect judging from the
silicate feature (its depth implies $A_{\rm V} \ge 15$ mag).
Therefore, either hard radiation leaks out because of clumps or
funnels created by supernova explosions, or there are stars in M82
outside the opaque nuclear dust clouds.  As our model cannot handle
clumping, but we nevertheless wish to extend the spectrum into the UV,
we simply add another stellar component.  It is not included in a
self--consistent way, but as its luminosity is $\sim 10\%$ of the
total, such an approximation may be tolerable.  The stellar
temperature and foreground reddening of the additional component are
poorly constrained (see caption of Fig.\ref{M82.ps}).  This is also
reflected by the controversial interpretations via an old stellar
population (Silva et al. ~1998) or via young, but obscured stars
(Efstathiou et al.~2000).

\noindent 
We mention that Sturm et al.~(2000) contest the existence of the 10$\mu$m
silicate absorption feature in M82.  They think that the value $\tau(18\mu$m) /
$\tau(9.7\mu$m) is too low and that recombination line ratios, like
H$_\beta$/H$_\alpha$, indicate only $A_{\rm V} \simeq 5$.  However, the small
ratio $\tau(18\mu$m)/$\tau(9.7\mu$m) is a radiative transfer effect where
18$\mu$m emission is favored over 10$\mu$m emission, and the
H$_\beta$/H$_\alpha$ ratio increases when the dust is not a foreground screen,
as Sturm et al.~assume, but mixed with the HII gas.  The strongest argument that
the 10$\mu$m depression is due to silicates comes from the 1mm flux.  It implies
a large column density of cold dust which, unless it is all behind the hot dust,
must produce the absorption feature. \\

{\bf {\it NGC253 }}
\smallskip

\noindent 
NGC253, another bright and nearby starburst, shows at high resolution
in the MIR a complex structure with several knots (Galliano et
al.~2005).  Nevertheless, the low spatial resolution observations are
well reproduced in our fit (Fig.\ref{sed1.ps}) which, in the $10 -
40\mu$m wavelength range, is of similar quality as in M82
(Fig.\ref{M82.ps}).  Below 2 Jy, ISOSWS data are noisy and have
therefore been omitted.  The dip at 18$\mu$m in the model of Piovan et
al.~(2006) is not present in ours, and not borne out by the
observations.  As our model dust is uncoated, the ice features
reported by Imanishi et al.~(2003) are not reproduced.\\

{\bf {\it NGC7714}}
\smallskip

\noindent 
Spitzer spectra (Brandl et al. 2004) do not reveal AGN signatures and support
our interpretation that the nucleus is dominated by a weakly obscured starburst
(Fig.\ref{sed1.ps}). \\

{\bf {\it NGC1808}}
\smallskip

\noindent 
This starburst is claimed to be young (Krabbe et al.~1994).  In high resolution
MIR images, several hot spots are detected which coincide with the most intense
radio sources (Galliano et al. 2005).  Our fit (Fig.\ref{sed2.ps}) somewhat
underestimates the PAH emission.  So one may have to increase the PAH abundance
(as was done in the model of Siebenmorgen et al.~2001) which is set constant in
the computations of our SED library.  Piovan et al.~(2006) predict silicate
absorption features at 10 and 18$\mu$m which, however, are not detected.  The
dip at $\sim 10\mu$m is due to the wings of neighboring PAH bands.  It is not
caused by silicate self--absorption which would require much higher optical
depths. \\

{\bf {\it NGC7552}}
\smallskip

\noindent 
This infrared luminous galaxy harbors a ring--like circumnuclear
starburst (Siebenmorgen et al.~2004).  Neglecting such structural
details, our fit to the dust emission is satisfactory. Two models are
shown in Fig.~\ref{sed2.ps} which bracket available data. The hot spot
density is low and the OB luminosity ratio, $L_{\rm OB}/L_{\rm tot}$,
is not well constrained (Tab.~1). \\

{\bf {\it NGC6240}}
\smallskip

\noindent 
NGC6240 is a merging ULIRG.  Such objects are one to two orders of
magnitude brighter than starbursts.  Our fit in Fig.\ref{ulirg.ps} is
acceptable despite a $\sim 30\%$ deficiency near 40--50$\mu$m.  To
better match the NIR photometry, we added to the starburst SED a 4000K
black body with $L = 10^{8.8}$\Lsun. Lutz et al.~(2003) suggest that
stars account for most ($\sim$75\%) of the total luminosity and that
the rest is due to an optically obscured AGN.  When they subtract from
the SED of NGC6240 a scaled--up M82 template, a faint (0.07Jy) residue
remains which they attribute to the AGN.  Dopita et al. (2005)
underestimate in their model the 10--30$\mu$m region (by a factor
$\sim$4 at 15$\mu$m) but they argue that they could match the data if
they added an AGN component, a procedure which is sometimes applied to
galaxies with hidden broad line regions (Efstathiou \& Siebenmorgen
2005). \\

{\bf {\it Arp220}}
\smallskip

\noindent 
Arp220 is the nearest example of a ULIRG. MIR high resolution maps
(Soifer et. al 2002) show a double nucleus with 1$''$ (360pc)
separation.  We process low resolution Spitzer IRS data using the SST
pipeline (Higdon et al. 2004).  The ISOPHT (Spoon et al.~2004) and
Spitzer spectrum reveals a complex spectrum with ice and silicate
absorption and pronounced PAH emission bands at 6.2 and 7.7$\mu$m.
Dopita's et al.~(2005) model predicts PAH features that are too strong
(factor $>5$).  Piovan et al.~(2006) fit the SED of the central 2kpc
region using an optical depth of $\tau_{\rm V}=35$mag and a dust model
with an SMC extinction curve.  Siebenmorgen et al. (1999) proposed
$\tau_{\rm V}=54$mag and MW dust to fit the photometric data available
at that time.  The SED library fit gives $R=3$kpc and $A_{\rm
V}=$72mag.  A model SED with $A_{\rm V}=$120mag and $R=1$kpc yields
too strong silicate absorption and requires an additional cold dust
component for the submm.

\begin{table}
\label{para.tab}
\caption{Fit parameters to models of Figs.~2--5}
\begin{center}
\begin{tabular}{l c c  c  c  c c }
\hline
                 &      &       &   &   &   &\\
Name             & $L^{\rm{tot}}$ & $D$& $R$ & $A_{\rm V}$ & $L_{\rm {OB}}/L^{\rm {tot}}$ & $n^{\rm {hs}}$ \\
                 & \Lsun  & Mpc  &  kpc  & mag & & cm$^{-3}$ \\
                 &      &       &   &   &   &\\
\hline
                 &      &       &   &   &   &\\
M82       & 10$^{10.5}$ & 3.5  & 0.35 & 36  & 0.4  & $10^4$  \\
NGC253    & 10$^{10.1}$ & 2.5  & 0.35  & 72  & 0.4  & 7500  \\
NGC7714   & 10$^{10.7}$ & 36.9  & 3  & 2  & 0.6  & 2500  \\
NGC1808   & 10$^{10.7}$ & 11.1  & 3  & 5  & 0.4  & 1000 \\
NGC7552   & 10$^{11.1}$ & 22.3  & 3  & 7  & 0.6  & $100$  \\
{\/ \/ \/} $''$    &  $''$   & $''$     &   $''$   & 9  & 0.4  &  $''$    \\
\hline
                 &      &       &   &   &   &\\
NGC6240   & 10$^{11.9}$ & 106  & 3  & 36  & 0.6  & $10^4$  \\
Arp220    & 10$^{12.1}$ & 73 & 1  & 120  & 0.4  & $10^4$  \\
{\/ \/ \/} $''$    &   $''$   &   $''$    & 3  & 72  &  $''$    &  $''$    \\
\hline
\end{tabular}
\end{center}
\end{table}

   \begin{figure}
\hspace{-0.5cm}{\includegraphics[angle=90,width=9.5cm]{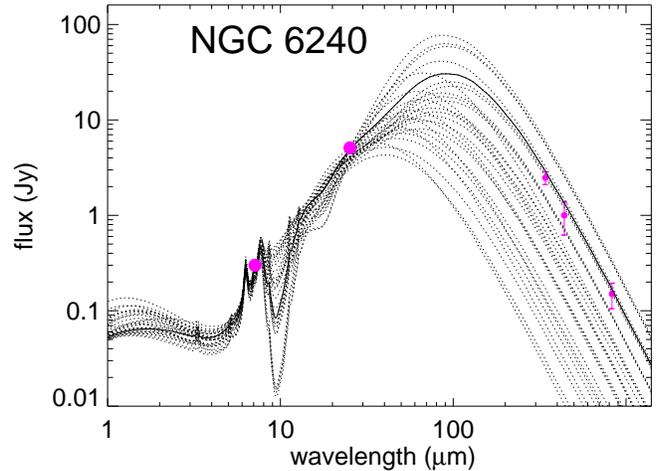}}
   \caption{All elements of the SED library (dotted) which fit 8 and 24$\mu$m
photometry (circles) of NGC6240 to within 30\%. Best fit (full line)
and other data as of Fig.\ref{ulirg.ps}. \label{2dat.ps}} 
\end{figure}

\subsection{Predicting fluxes}

\noindent 
The infrared luminosity is the key parameter of a galaxy and it is
often used to estimate the star formation rate of a galaxy (Kennicut
1998). Unfortunately, for faint or redshifted objects photometry is
sometimes only provided at two MIR bands, for example at 8 and
24$\mu$m from the Spitzer satellite.

\noindent 
In Fig.~\ref{2dat.ps} we demonstrate for the ULIRG NGC6240 that the
SED library can be used to estimate the total luminosity to within a
factor $\sim 2$ from two such MIR fluxes only.  We also see
(Fig.~\ref{2dat.ps}) that the SED will be quite well constrained by an
additional submm data point.

\section{Discussion}

\subsection{Methods of modeling starburst SEDs}

\noindent 
One finds in the literature three different ways to reproduce or explain the SED
of an extragalactic object.

{\it i)} Matching it with a template SED of a well known galaxy
(Laurent et al 2000, Lutz et al. 2003, Spoon et al. 2005).  This is
reasonable only as long as template and object are similar in their
parameters as well as geometrical structure and orientation on the
sky: So it is not meaningful to compare SEDs of AGN type 1 and type 2,
or objects with radically different luminosities, like M82 and
NGC6240, because the luminosity affects the SED, as can be seen from
Fig.~\ref{para4.ps}a.  To obtain a good match between the SED of the
object and the template, after normalization to a unit distance, one
usually has to scale the flux of the template moderately up or down to
fine--tune the luminosity.  If the fit is successful one may argue
that the object is similar to the template in its geometry and basic
parameters (but for some scale factor) and that one understands it
almost as well as the template.

{\it ii)} Reproducing the shape of the SED by optically thin dust
emission.  The dust is assumed to be heated in a given radiation field
which is usually a scaled--up version of the interstellar field
(Devriendt et al. 1999, Dale et al. 2001, 2005, Lagache et al. 2003).
This procedure neglects all effects of radiative transfer and must
fail when dust self--absorption becomes important, most strikingly in
the $\sim 10\mu$m region as shown in Fig.~\ref{para4.ps}d.

{\it iii)} Solving the radiative transfer, in various degree of sophistication.
This is a much more ambitious method and requires assumptions about the
structure and parameters of the galaxy.  Three--dimensional codes have been
applied to spirals (Kylafis \& Bahcall (1987), Popescu et al.~(2000), Tuffs et
al.~(2004)) using ray tracing or Monte Carlo techniques (Bianchi et al.~2000).
Rowan-Robinson \& Crawford (1989) fit IRAS color diagrams of starbursts using a
one--dimensional transfer code.

\medskip
\noindent 
We also do radiative transfer calculations and it may be instructive
to point out technical and conceptual differences between our models
and those devised recently by other authors (Silva et al. (1998),
Efstathiou et al.~(2000), Takagi et al.~(2003), Dopita et al.~(2005)
and Piovan et al.~(2006)), although we admit that we did not always
find it easy to pin down exactly which approximations our colleagues
used (as they may experience difficulties in identifying our
assumptions).

\noindent 
All groups evaluate the emission from a dusty medium of spheroidal shape filled
with stars, all seem to use similar optical dust constants, and all incorporate
small grains with temperature fluctuations (like PAHs).  At first glance, the
model results appear to agree, but upon closer inspection one finds that the
maximum deviations in the other papers are considerable (factor 4), whereas the
fits from our library to the prototype objects (Fig.2 to 5) are nowhere off by
more than 50\%.

\medskip
\noindent 
The major points where our paper differs concern the treatment of the sources,
the interstellar extinction curve, the radiative transfer, and the presence or
neglect of hot spots.

{\it a) Stellar sources.}  We do not take into account the time
evolution of a stellar population after the burst and the possibility
that there may have been several episodes of rapid star formation.
Our models are therefore simpler and do not allow to constrain the age
of the burst(s).  We are, however, skeptical that this is possible
without spectroscopic observations (like stellar CO bands in
supergiants), also derived ages of the stellar populations are
controversial (see M82).

{\it b) Extinction curve.}  We assume galactic dust and do not consider the
possibility that it may be a combination of the species found in the Milky Way,
LMC or SMC.  Again, here our model is simpler, but as the extinction of the
sources is usually large ($A_{\rm V} > 5$ mag), the exact shape of the reddening
curve has little effect on the resulting infrared SED.

{\it c) Radiative transfer.}  As far as we can tell, the interaction
between dust and radiation is treated consistently only in Takagi et
al.~(2003) and the present paper.  Other authors introduce basic
emission units which they compute separately.  Such a unit may be a
diffuse gas cloud, or a spherical molecular cloud filled with dust and
stars that are either continuously smeared out over the cloud or
concentrated in the center.  The emission units are then scaled up, or
a simplified radiative transfer (with constant source function or no
reemission) is applied to match the nucleus under consideration.
Naturally, when the optical thickness is not small, models without
radiative transfer are at some point faulty, although it is hard to
quantify how much the simplifications effect the resulting SED.

{\it d) Hot spots.}  They are a particular feature of our models and
inevitably arise when a luminous star is enveloped by a cloud with a
density considerably above the mean density of the nucleus.
Neglecting hot spots seriously underestimates the MIR emission of the
nucleus (see Kr\"ugel \& Siebenmorgen (1994) or Fig.~\ref{para4.ps}c).

{\it e) Clumpiness.} As discussed in the model description of M82, the
optical and UV flux is best explained by postulating the interstellar
medium to be clumped.  Clumping is a natural consequence of SN
explosions; when the surface filling factor is close to one, it has
little effect on the SED at wavelengths greater than a few micron.
The models that include stellar evolution (Silva et al., Efstathiou et
al., Takagi et al., Dopita et al., Piovan et al.) introduce as an
additional free parameter the fraction of starlight escaping the
galaxy due to clumping; this fraction depends in their computations on
the age of the starburst.  Our approach is again simpler.  Because the
UV and optical stellar light that leaks out is in reality modified in
a complicated way by the passage through a clumped medium, we only add,
where necessary, a blackbody curve to account for the excess light.

\subsection{Completeness, uniqueness and credibility of the SED library}

It is remarkable that one can very well fit the SEDs of galaxies, like M82,
Arp220 and others, with models of constant density and radial symmetry.  The
satisfactory fits imply, first, that our library grid is sufficiently fine, and
we expect that starbursts observed with similar wavelength coverage as those
presented in Fig.2-5 (more than a few data points in the SED) can be reasonably
matched by a single element of the SED library.

\noindent 
Nevertheless, one may wonder whether the fits are meaningful.  After
all, we know for AGN, which have tori that lead to the division into
type 1 and 2 with respect to the observer, that spherical symmetry is
a principally unacceptable approximation.  As the torus is the result
of rotation, it should form independently of a massive black hole and
therefore also exist in starbursts.  However, there seems to be no
need to invoke one. There are probably two explanations.  First,
whereas an AGN is small (pc) and easily shadowed by the much bigger
torus (100 pc), a starburst region is as large as or larger than a
torus and could not be blocked visually.  So there cannot be
starbursts of type 1 and 2.  Second, the galaxy collision preceding
the starburst leads to strong perturbations of the nuclear gas which,
in the gravitational potential of the little disturbed bulge stars,
results in rough spherical symmetry.

\noindent 
We also have to discuss the possible contamination of starburst fluxes by
emission from the galactic disk when the spatial resolution of the observations
is poor.  This is the standard situation in the far infrared.  There, however,
the nucleus is usually much brighter than the disk and then the contamination is
irrelevant.  It may be substantial at short wavelengths (NIR, optical, UV) if
the starburst is very obscured and little optical flux leaks out.  

\noindent 
In the submm/mm region, one measures mainly the dust mass and there is likely
to be more mass in the disk than in the core.  To estimate the
contributions of the disk and the core, let $F, L, M$ and $T$ denote
the observed flux, bolometric luminosity, dust mass and dust
temperature, respectively.  With the approximations $L\propto M T^6$
and $F_{\rm 1mm} \propto M B_{\rm 1mm}(T) \propto M T$, the flux
ratio, say, at 1 mm, becomes $F_{\rm 1mm,c}/F_{\rm 1mm,d} = (L_{\rm
c}/L_{\rm d}) (T_{\rm d}/T_{\rm c})^5$.  Here we used the index {\it
d} for disk and {\it c} for the core.  Typical mean values are $T_{\rm
d}=10 \ldots 20$K (Kr\"ugel et al. 1998) and $T_{\rm c}= 30\ldots 50$
K (Klaas et al. 2001).  Therefore, the cold dust in the disk is not
really important in measurements of low spatial resolution as long as
the nucleus is much brighter than the disk.

\section{Conclusions}

We have computed in a self--consistent radiative transfer SEDs of spherical,
dusty galactic nuclei over a wide range of their basic parameters such as
luminosity, dust mass, size and obscuration.  The SEDs can be accessed in a
public library \footnote{ The SED library is available at: \\{\tt
http://www.eso.org/\~ \/rsiebenm/sb\_models}}.

\noindent 
Given a set of data points for a particular galaxy, there is a simple
procedure, described in the README file, to select from the library
those elements which best match them.  If the observations cover the
full wavelengths band from a few $\mu$m to about 1 mm, one usually
finds only one library element that fits very well, as demonstrated
for seven famous active galaxies.  If the data points are widely
spaced, there may be a few elements of less fitting quality, but
similar in their basic parameters.

\noindent 
The library therefore allows one to constrain the fundamental
properties of any nucleus which is powered by star formation and for
which data exist, without any further modeling.  Two observed fluxes
in the MIR plus one submm point are usually sufficient for a crude
characterization of the nucleus.  If there are only two MIR points,
for example, from Spitzer at 8 and 24$\mu$m, one can still bracket the
total luminosity within a factor of $\sim 2$.

\noindent 
In the UV, optical and NIR, it may be necessary to add to the library SED a low
luminosity stellar component, at least, this was necessary for M82 and NGC6240.
This component may be due to photons that escaped the nucleus without
interaction because of clumping or it may be light from the galactic disk.  

%We argue that excess emission in the submm due to very cold dust is not
%important as long as the luminosity of the burst dominate.

\begin{acknowledgements}
      We thank D. Lutz for providing the observed SED of NGC6240 and
      V. Charmandaris for the Spitzer IRS spectrum of NGC7714. This
      research has made use of the NASA/IPAC Extragalactic Database
      (NED) which is operated by the Jet Propulsion Laboratory,
      California Institute of Technology, under contract with the
      National Aeronautics and Space Administration.

\end{acknowledgements}

\end{document}